# Review: Halogens in the synthesis of colloidal semiconductor nanocrystals

*Frauke Gerdes,[1] Eugen Klein,[1] Sascha Kull,[1] Mohammad Mehdi Ramin Moayed,[1] Rostyslav Lesyuk,[1,2] Christian Klinke[1,3]*

[1] Institute of Physical Chemistry, University of Hamburg,
Grindelallee 117, 20146 Hamburg, Germany

[2] Pidstryhach Institute for applied problems of mechanics and mathematics of NAS of Ukraine,
Naukowa str. 3b, 79060 Lviv, Ukraine

[3] Department of Chemistry, Swansea University - Singleton Park,
Swansea SA2 8PP, UK

**Abstract**

*In this review, we highlight the role of halogenated compounds in the colloidal synthesis of nanostructured semiconductors. Halogen-containing metallic salts used as precursors and halogenated hydrocarbons used as ligands allow stabilizing different shapes and crystal phases, and enable the formation of colloidal systems with different dimensionality. We summarize recent reports on the tremendous influence of these compounds on the physical properties of nanocrystals, like field-effect mobility and solar cell performance and outline main analytical methods for the nanocrystal surface control.*

**Introduction**

Since the hot injection method was implemented to produce high-quality, highly monodisperse, spherical chalcogenide nanocrystals (NCs) [1], many approaches have been developed to control NC shape, crystal phase and dimensionality. During the last decades the use of highly reactive organometallic compounds as metal precursors has been replaced by greener compounds including halogen-containing metal salts [2-5]. These compounds allow obtaining different NC shapes ranging from dots, to rods and tetrapods [5-8]. In these syntheses, the halide anions play relevant roles in both the nucleation and growth stages.

Apart from the direct use of halogen-containing metal salts used as metal precursors, halide ions can be released by additional halogenated hydrocarbons included in the synthesis (like dichloroethane, dibromoethane, and diiodoethane), as additional reagents [9, 10]. In particular, haloalkanes can operate as molecular ligands or decompose to halide ions while passivating specific facets in order to stop the growth or to increase the reactivity of these facets [6]. Halogen compounds are also able to modify the shape of NC during the synthesis by ripening processes [6].

An extensive review by Juarez [11] summarizes the influence of chlorinated hydrocarbons on the shape and surface chemistry of wurtzite CdSe NCs. Comprehensive studies using X-ray photoelectron spectroscopy (XPS), mass spectrometry (MS), nuclear magnetic resonance (NMR) spectroscopy, and cyclic voltammetry measurements clarified the mechanism of chloride incorporation into the ligand shell and concomitant NCs shape modification. The role of chloride ions altering the kinetics and influencing the shape evolution of NCs during growth has been investigated in the last years more intensively [9, 10, 12-18]. These studies highlight that halogens induce a strong influence on the surface chemistry of NCs. In a different approach, post-synthetic treatments of colloidal NCs with halogenated compounds can lead to the stabilization of a particular morphology or can improve the electrical properties as a result of doping or better surface passivation [13-16]. In the following, the role of halogenated compounds on the control of shape, crystal phase and improvement of electrical and optical properties will be discussed.

**Chemical and physical background for the ligation of NCs with halogens**

To saturate the dangling bonds on the NC surface appropriate ligation is needed. The most commonly used ligands belong to the groups of neutral L-type or ionic X-type ligands [19, 20]. The L-type ligands (neutral donor) like amines, carboxylic acids or phosphonic acids can coordinate both cationic and anionic surface atoms, while the X-type ligands (ionic donor of an electron) like halide ions, carboxylates or thiolates selectively bond to positively charged atoms [21]. If both types are present, the X-type ligands will dominate on the surface due to the stronger interaction with the surface. This feature of halogens was recently supported by comparison of the adsorption energy of different ligands on the basis of calculations in the frame of density-functional theory (DFT) [22]. Halide ions belong to the group of X-type ligands, though there are differences between them. The nucleophilicity and the ability being a good leaving group increase in the following order $F^- < Cl^- < Br^- < I^-$. As a result, the release of the halide ions from single haloalkanes is easier for the heavier $I^-$ followed by $Br^-$, $Cl^-$ and $I^-$. This means that for a particular reaction at a fixed temperature, haloalkane concentration and reaction time, the released amount of $I^-$ will be higher than that of $Br^-$ followed by $Cl^-$ and $F^-$ [11]. The presence in larger amount of halide ions at earlier stages of the synthesis will in turn modify the nucleation step and growth process and can lead to shape and crystal phase transformations. The nucleophilicity tendency of the halide ions also affects the reactivity of haloalkanes. Moreover, the adsorption energy of the halide ions decreases the heavier the atom. Fluoride and chloride ions usually have different influence on the NC shape than bromide and iodide ions due to the stronger coordination to the metal surface atoms.

Beside the type of ligand, the hard-soft acid-base (HSAB) concept has also to be taken into account. In theory, soft acids react faster and form stronger bonds with soft bases, whereas hard acids react faster and form stronger bonds with hard bases [23]. The hardness of the halide ions decreases with heavier atom due to the increased radius and due to the decreased electronegativity. This correlation for halogens as X-type ligands was successfully verified by DFT simulations on the example of CdSe, ZnSe and InP NCs [22]. Depending on the hardness or softness of the metal as well as the other used ligands, halide ions are more effective und successful to influence the final NC [24]. For example, $Cu^+$ is a soft acid and favors soft base ligands like iodide and thiols, whereas $Fe^{3+}$ is a hard acid and prefers hard base ligands like fluoride and amines.

**Morphology control**

Since the mid-2000s, it has been shown that the addition of halogens as ligands plays a prominent role in the formation of different shapes of NCs, leading to structures like sheets, pyramids or rods [9, 25]. In 2007, Yong with co-workers showed the preparation of a variety of differently shaped CdS NCs including nanorods, bipods, tripods, tetrapods, nanocubes, nanowires, and nanoplatelets [17]. The syntheses of these materials were performed at temperatures between 100 and 175 °C and consisted of the reaction of a metallic salt dissolved in oleylamine with an oleylamine sulfur precursor and the corresponding co-ligands such as cetyltrimethylammonium bromide (CTAB) or hydrochloric acid (HCl). When CTAB was used as a co-ligand, CdS chain-like NCs were observed as shown in Figure 1 A. These structures have a strong tendency to stack and it was suggested that the change in shape, compared to NCs synthesized without co-ligands seen in Figure 1 B and C, occurs due to an oriented attachment mechanism triggered by the passivation of different facets on the CdS NCs surface. In the case of HCl as co-ligand the product obtained was CdS nanorods, which showed crystals with higher uniformity compared to rods prepared by other methods.

In addition to CdS, various other materials such as CdSe, PbS, CuS and $Cu_{2-x}S$ have been synthesized in different shapes using halide ligands [18, 24-28]. A shape transformation from CuS nanodisks to triangular nanoprisms was reported [18] when chloride, bromide and iodide ions were used as co-ligands (Figure 2 A,B). Donega et al. reported a decisive role of bromide regarding the morphology of copper sulfide (tetragonal digenite, $Cu_{2-x}S$) nanosheets (NSs) by using tin bromide as additive in the synthesis. The presence of Br in the product was revealed by EDX spectroscopy. By increasing the amount of $SnBr_4$ in the flask the authors were able to switch the product from small NCs to nanodiscs and micron-sized NSs with unprecedented thickness of only 2 nm [24].

For PbS NCs Cademartiri et al. presented a synthesis based on a $PbCl_2$-oleylamine (OLA) complex forming a highly viscous reaction medium [29]. By XPS analysis the authors found that beside oleylamine, chloride ions are present on the NC surface and adopted the role of ligands. Other authors synthesized PbS, ZnS, CdS, and MnS NCs using metal chlorides in oleylamine, although in these cases the role of the remaining chloride ions in the synthesis was not discussed [12].

CdSe NCs were extensively studied to control shape and crystal phase by halogenated compounds. In a previous work we observed a shape transformation from hot-injection prepared CdSe nanorods to hexagonal dipyramidal NCs during the preparation of composites with carbon nanotubes, which were dispersed in 1,2-dichloroethane [26]. To understand the occurred shape evolution we studied the reaction in a modified synthesis without carbon nanotubes but with various halogen compounds [9]. It was found that the CdSe pyramids were formed through ripening of rod-shaped NCs by released chloride, bromide and iodide ions from haloalkanes. The tendency to release these ions and consequently to increase the concentration during the synthesis was important for the final shape and size of the CdSe pyramids. Higher concentrations of halide ions in the beginning of the synthesis lead to larger rods and pyramids. XPS and total reflection X-ray fluorescence measurements showed that the shape transformation is triggered by the replacement of phosphonate ligands and the incorporation of halide ions on the surface.

The growing interest in two-dimensional (2D) materials has triggered synthetic efforts also in colloidal synthesis. Colloidal 2D NCs can be synthesized by several means including the formation of lamellar structures as a soft template [30], the oriented attachment of 0D dots in XY plane restricting the Z direction [25] and by assembly of tiny clusters to nanoplatelets with atomically precise thickness [31]. Chloroalkanes were succesfully used for the preparation of sheets of PbS in 2010 reported by Schliehe et al. where the {110} facets were covered by 1,1,2-trichloroethane. This increased the reactivity of the mentioned facets leading to a specific arrangement of the NCs and resulted eventually in ultrathin sheet structures [25] changing the dimensionality of the NC from zero-dimensional (0D) to two-dimensional (2D) through the oriented-attachment mechanism [32-34]. Deeper investigations on the role of chloride ions in the colloidal PbS synthesis revealed the possibility to control the shape of the product from dots to stripes with improved electronic properties and further to NSs by the amount of chloride in the system (discussed later in this review text) [27]. Sun et al. demonstrated the NSs thickness in the range of 2 – 5 nm depends on the chloroalkane used during the reaction [28].

**Crystal phase control in II-VI system**

Until now, we showed the importance of halide ions during the growth and ripening phase to control the shape of NCs. However, halogen compounds can also play a critical role in the determination of the crystal phase of NCs. In 2010, Saruyama et al. showed the crystal phase transformation from zinc blende CdS NCs to wurtzite CdS "nanopencils" [7]. The combination of chloride ions and surfactants (oleic acid and oleylamine) led to a dissolution of the NCs and the following growth of larger ones, whereby the size of the nanopencils dependents on the amount of chloride in the reaction solution. In contrast, Tai et al. showed that the crystal phase transformation from zinc blende to wurtzite CdS NCs occurred with increasing NaCl concentration in an ultrasound-assisted microwave synthesis without further ligands [35]. They suggested that chloride ions act as a capping agent and inhibited the growth of the CdS (111) surface, while the growth of the (100) and (101) planes is favored. Lim et al. also found that the halide ion density on the surface is responsible for the formation of wurtzite "arms" on zinc blende CdSe seeds [6]. A quantitative $^1$H-NMR study showed that the amount of oleate decreased with increasing halide ion content and the anisotropic growth is induced by the replacement of surface oleate ligands and the destabilization of the CdSe NCs. Zou et al. also observed that the exchange of oleate through halide ions in the heat-up synthesis of CdS NCs leads to a crystal phase change from zinc blende to wurtzite [8]. In contrast to the above-mentioned synthesis, they proposed that control over shape and crystal phase takes place during the early stages of the synthesis, not during the growth regime. According to the HSAB concept, they suggested that the soft base halide ions bind to the soft acid cadmium and replace the hard base oleate. This strong binding leads to a reduction of the precursor reactivity, decreasing the amount of nuclei during the nucleation, and favoring the formation of larger NCs. Earlier studies have proven that the crystal phase of CdS NCs is size-dependent [36, 37].

Beside the mentioned crystal phase transformation and change for spherical-shaped and complex-shaped CdS and CdSe NCs by halide salts, the phase control is also possible for 2D CdSe NCs by using haloalkanes [38]. With increasing amount of 1-bromoheptane in a hot-injection system the shape of the ultrathin CdSe NSs changes from sexangular (Figure 3 A) to quadrangular (Figure 3 B–D) to a mixture of different shapes (Figure 3 E) to triangular (Figure 3 F,G) ones. The X-ray powder diffraction (XRD) patterns (Figure 4) indicate that a change in the crystal phase from zinc blende to wurtzite occurred as the shape of the CdSe NSs varied. The observation of bulk $CdBr_2$ in the XRD pattern at high amounts of 1-bromoheptane suggest that bromide ions must be formed *in situ* and that they might be responsible for the shape and crystal phase change of the CdSe NSs. To determine the role of the bromoalkane in the reaction electron ionization MS was performed, where samples were obtained before and after the injection of 1-bromoheptane. The mass spectrum reveals the presence of cadmium acetate [(CH$_3$COO)-Cd-(OOCCH$_3$)] and a cadmium acetate bromide complex [Br–Cd-(OOCCH$_3$)], evidencing the replacement of acetate through the haloalkane in the reaction due to the HSAB concept. To determine the active species binding to the CdSe NS surface XPS were performed. These measurements showed that the Br 3d area can be fitted to covalently bonded bromine atoms in Br-Hep and to ionic bonded bromine atoms, indicating different chemical Br environments. The observed shape change in the CdSe 2D nanostructures is explained on the basis of the bromide ions, which partially replace the carboxylate ligands on the surface and modulate the surface energy of the facets during the growth regime. When the concentration of Br-Hep and therefore the resulting amount of bromide ions is increased the crystal phase of the CdSe NSs also varies. In this case, the nucleation process is influenced by the presence of higher amounts of [Br–Cd-(OOCCH$_3$)]. Due to the strong Cd-Br bond the activated complex formed upon the reaction with Se-P(octyl)$_3$ is less reactive than the previously evidenced general complex based on the cadmium acetate complex. As a result of the reduced precursor reactivity the P=Se cleavage rate is slowed down and the nucleation rate is reduced. In this scenario, the crystal phase of the CdSe nanostructures is wurtzite. The use of other bromoalkanes also leads to triangular, wurtzite CdSe NSs but the minimum required amount of bromoalkane differs notably due to the different reactivity to release bromide ions.

**Influence of halogens on physical properties of colloidal nanocrystals**

The attractiveness of the above-described effects (dimensionality, shape and crystal phase control with halogens) is highly supported by the fact that halide ions are able to alter and/or improve the physical properties of colloidal NCs. Halide ions can be provided to the crystals in different stages of the synthesis, resulting in better surface passivation or doping, which can eventually lead to an enhanced efficiency of devices produced with such NCs [14, 39-46].

Usually, the surface of colloidal NCs shows a high amount of trap states. The mismatch between the periodicity of the crystal and the coverage of the surface ligands leads to the formation of dangling bonds and consequently to surface trap states [43, 47]. Several effects

-7-

are anticipated from these trap states. First, the carrier concentration is reduced since charge carriers are trapped in these states and cannot contribute to the transport anymore [43, 47, 48]. Further, by scattering the mobile carriers, the charge mobility in the crystal is reduced [43, 49-51]. Finally, by altering the band structure, mid-gap states are formed, functioning as recombination centers. Mid-gap states dramatically worsen the gateability of the crystals, their conversion efficiency (as solar cells), as well as their emission intensity [14, 40, 43, 44, 47, 49, 51, 52]. In order to address these issues, employing halide ions as X-type elemental inorganic ligands in addition to the organic ligands has been suggested (hybrid passivation) [14, 39-46].

Several strategies have been used to incorporate halide ions. For instance, PbS or PbSe quantum dots have been synthesized by using $PbCl_2$ or $PbBr_2$ as the precursor. In these systems, halide ions were obtained during the synthesis by the decomposition of the precursor, covering the surface of the quantum dots, which was confirmed by XPS measurements and EDX mapping [40, 46]. These NCs were able to maintain their absorbance at least for 35-40 days of exposure to air, since their passivated surface was strongly resistant against further oxidation (Figure 5 A) [40,9]. Also their photoluminescence quantum yield could be improved by up to 30%, while their power-conversion efficiency as solar cells could reach 6.5% [40] or 7% [14]. A similar strategy was employed to produce InP tetrahedral NCs (by using indium trichloride as precursor) [45]. Alternatively, as mentioned in previous sections, chloroalkanes can be added to the synthesis as co-solvents (co-ligands). Kandel et al. investigated the effect of chloride ions on the PbSe nanorods formation. Successful co-passivation by chloride ions was revealed by EDX spectroscopy. The photoluminescence quantum yield of these NCs has been reported to be about 27%, confirming well-passivated surfaces [53].

As a distinct approach (the so-called post-synthetic treatments) halide ions can be introduced to the NCs after the synthesis, when the product has been completely formed. For instance, an ammonium chloride solution (or other ammonium halides [39]) was added to the reaction mixture after cooling the solution down. This method has been used for passivation of PbS NCs [39] or PbSe NCs (with or without a PbS shell) [41] and has led to air stability of the particles for weeks. The absorption spectra of these particles remained unchanged after some weeks of air exposure, while without the treatment they experienced a fast degradation (Fig. 5 B) [39, 41]. Further, their functionality as field-effect transistors (FETs) has been preserved even in air, demonstrated by a pronounced n-type behavior [39]. In this respect, calculations based on DFT revealed that by increasing the atomic number of the halide, the formation energy of the $PbX_2$ adlayers (X: F, Cl, Br, I) becomes higher, resulting in higher probabilities for the $PbX_2$ formation and therefore, more effective passivation [39].

Eventually, halide ions can be employed to passivate the surface of the NCs separately from the synthesis and in a post-synthetic ligand exchange process, either in the solution state or in the solid state (on films of particles) [42, 48]. XPS measurements on these NCs showed that the solid state ligand exchange (halide treatment) was more effective, since in the solution phase, halide ions are more susceptible to the protic attack of the solvent, and therefore, prone to the desorption from the surface of the particles. This weakness became more pronounced by reducing the atomic number of the halide, consistent with the HSAB concept ($I^-$ → $Br^-$ → $Cl^-$) [42]. PbS NCs passivated by this method showed n-type behavior as well as stable properties in air. FETs based on these NCs exhibited n-type conduction with a field-effect mobility of 0.1 $cm^2V^{-1}s^{-1}$ and an on/off ratio of $10^5$ (Figure 6 A) [48]. As can be seen in Figure 6 B, the hybrid-passivated NCs show the best photovoltaic performance compared to the particles passivated only with organic ligands or only with inorganic ligands [14]. Their power-conversion efficiency could be increased to 8%, manifesting a great improvement compared to other quantum-dot based solar cells [42].

**Conclusion**

Understanding the role of halogens in the colloidal synthesis of semiconductor NCs allows establishing strategies to control the shape, dimensionality, crystal phase and physical properties. Halogens affect the surface chemistry of the NCs, playing either the role of a ligand or co-ligand passivating the surface, or influencing the growth and development of particular facets. Halogens can be included to the surface of NCs during the synthesis or during post-synthetic treatments such as ligand exchange in solution or in solid state. Tremendous improvement of physical properties such as field-effect mobility, stability, photoluminescence quantum yield and photovoltaic performance has been reported by NCs passivated by halides or doped with halogen giving a promising outlook for NC application in electronics and photovoltaics.

## ACKNOWLEDGEMENTS

The authors thank Beatriz H. Juarez for critically reading the manuscript and helpful suggestions. The authors gratefully acknowledge financial support of the European Research Council via the ERC Starting Grant "2D-SYNETRA" (Seventh Framework Program FP7, Project: 304980) and thank the German Research Foundation DFG for financial support in the frame of the Cluster of Excellence "Center of ultrafast imaging CUI". C.K. thanks the German Research Foundation DFG for financial support in the frame the Heisenberg scholarship KL 1453/9-2.

**Figures**

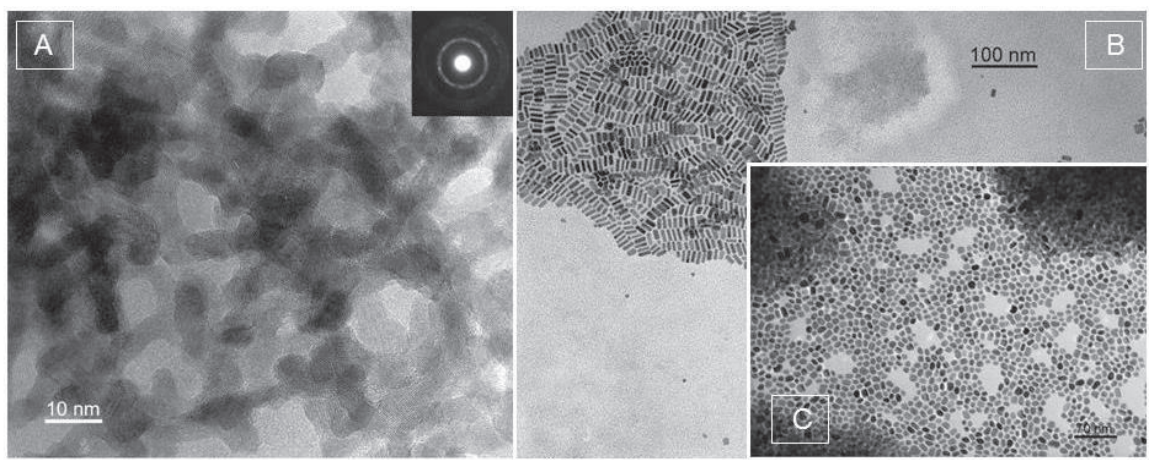

**Figure 1.** (A) High resolution TEM image of CdS chain-like NCs obtained with CTAB as co-ligand. (B) TEM image of uniform CdS nanorods synthesized with HCl. (C) TEM image of CdS NCs prepared without any co-ligand. [17]. Reprinted with permission from K. T. Yong, Y. Sahoo, M. T. Swihart, P. N. Prasad, J. Phys. Chem. C 111 (2007) 2447. Copyright 2007 American Chemical Society.

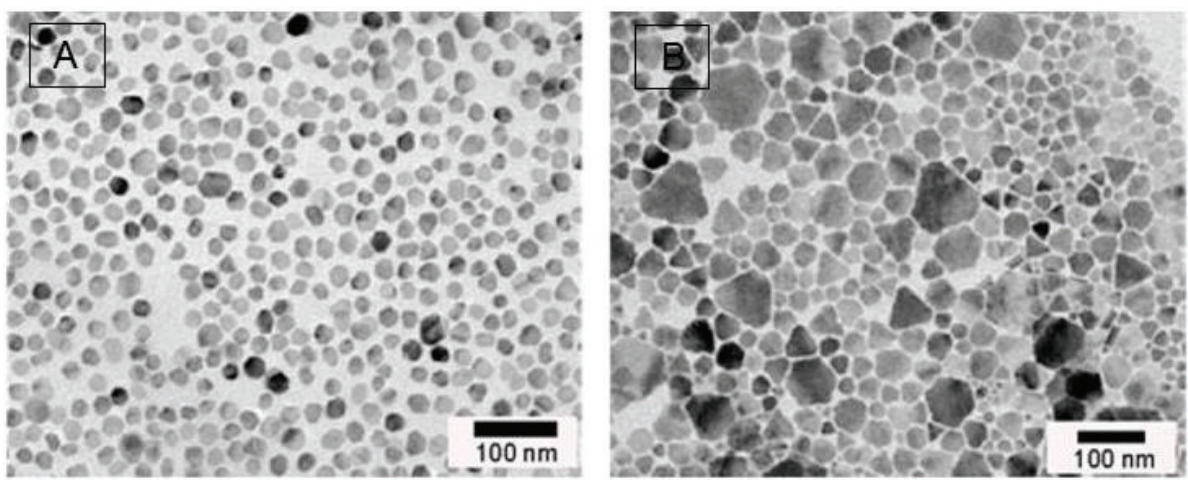

**Figure 2.** TEM images of CuS nanodisks without halide ions (A) and triangular nanoprisms synthesized with chloride ions (B) [18]. Reprinted with permission from S.-W. Hsu, C. Ngo, W. Bryks, A. R. Tao, Chem. Mater. 27 (2015) 4957. Copyright 2015 American Chemical Society.

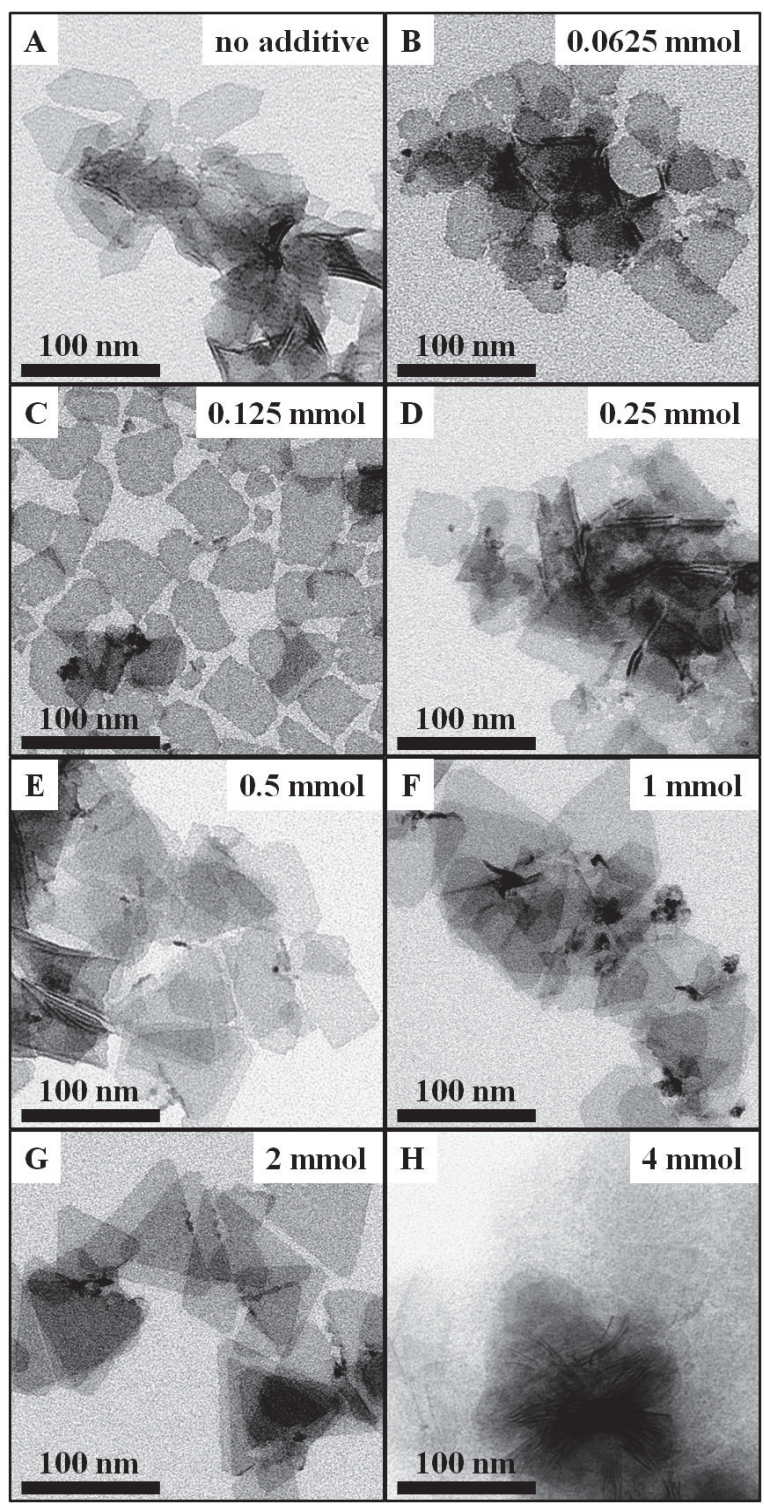

**Figure 3:** TEM images of ultrathin CdSe NSs synthesized with different amounts of Br-Hep. (A) Sexangular (no additive), (B–D) quadrangular (0.0625, 0.125, and 0.25 mmol Br-Hep), (E) mixture shaped (0.5 mmol Br-Hep), (F) triangular (1 mmol Br-Hep), (G) triangular (2 mmol Br-Hep), and (H) undefined NSs (4 mmol Br-Hep) [38]. Reprinted with permission from F. Gerdes, C. Navío, B. H. Juárez, C. Klinke, Nano Lett. 17 (2017) 4165. Copyright 2017 American Chemical Society.

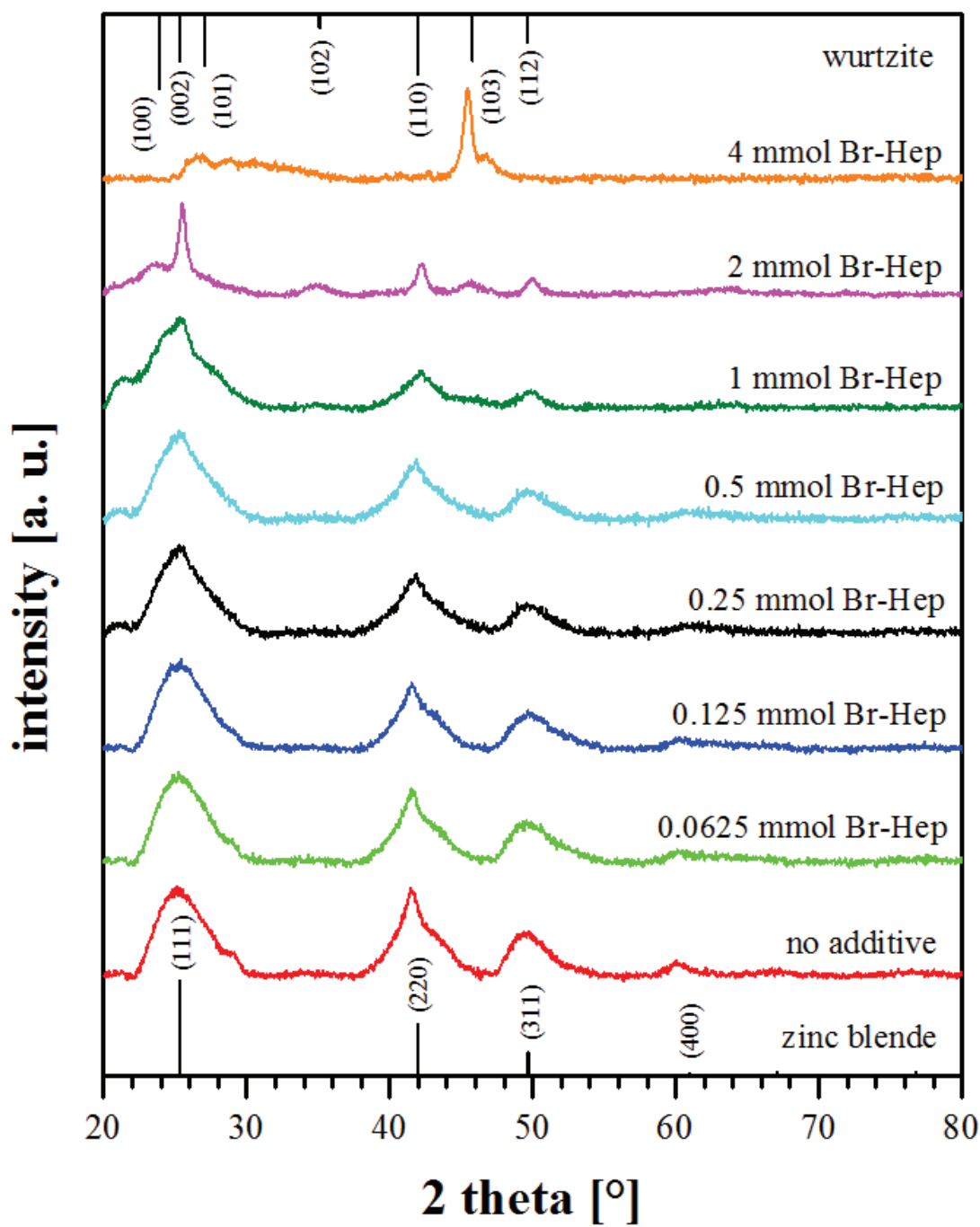

**Figure 4:** XRD of ultrathin CdSe NSs synthesized with different amounts of Br-Hep. At the top (wurtzite, ICPDS 00-008-0459) and the bottom (zinc blende, ICPDS 00-019-0191) the diffractograms of the bulk CdSe is shown. The peaks of the sample with 4 mmol Br-Hep show a strong texture effect and are assigned to the bulk $CdBr_2$ (ICPDS 00-010-0438) [38]. Reprinted with permission from F. Gerdes, C. Navío, B. H. Juárez, C. Klinke, Nano Lett. 17 (2017) 4165. Copyright 2017 American Chemical Society.

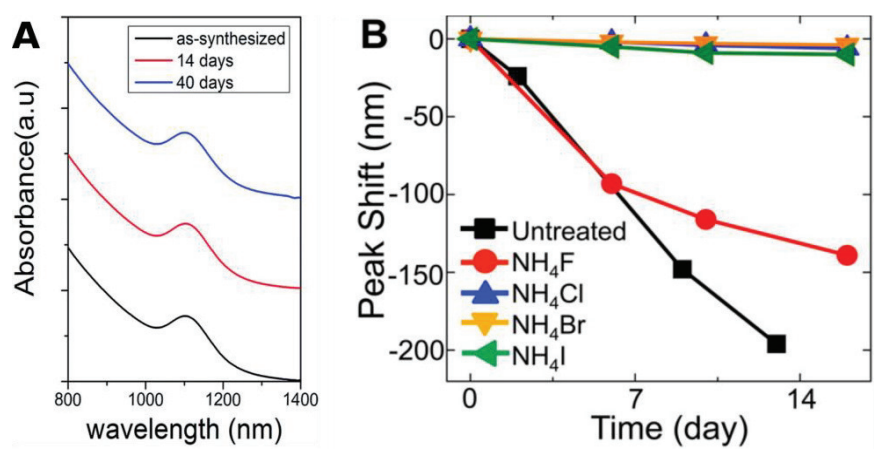

**Figure 5: Optical absorption properties of surface passivated NCs.** (A) Absorbance of PbS NCs passivated through the PbBr$_2$ precursor which shows air stability for at least 40 days of air exposure [46]. (B) Effect of different halides on the passivation of PbSe particles, added at the last stage of the synthesis after cooling down. Except fluoride, other halides had pronounced passivation effects [39]. Reprinted with permission from J. Y. Woo, J. H. Ko, J. H. Song, K. Kim, H. Choi, Y. H. Kim, D. C. Lee, S. Jeong, J. Am. Chem. Soc. 136 (2014) 8883. Copyright 2014 American Chemical Society.

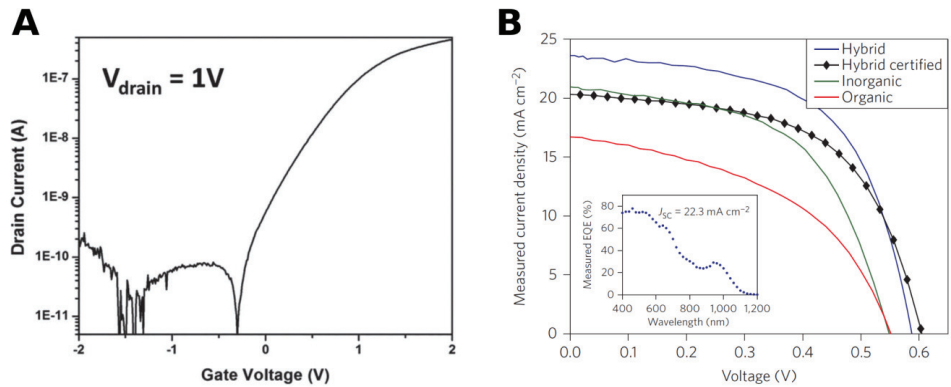

**Figure 6: Post-synthetic treatment of NCs with halide ions.** (A) Transistor behavior of PbS NCs passivated by iodide ions, showing pronounced n-type behavior with an on/off ratio of $10^5$ [48]. (B) Comparison between the effectiveness of different passivation methods on the I-V characteristics of the solar cells made of PbS particles, showing the best performance with hybrid passivation. Black diamonds are the I–V curve for a hybrid passivated device as measured by an accredited photovoltaic calibration laboratory (Newport Technology and Application Center–PV Lab). Inset: external quantum efficiency (EQE) curve for a hybrid passivated device. The integrated current value is also shown [14]. Reprinted by permission from Springer Customer Service Centre: Springer Nature, Nature Nanotechnology, A. H. Ip, S. M. Thon, S. Hoogland, O. Voznyy, D. Zhitomirsky, R. Debnath, L. Levina, L. R. Rollny, G. H. Carey, A. Fischer, K. W. Kemp, Nat. Nanotechnol. 7 (2012) 577. Copyright 2012.

-15-